\documentstyle[12pt]{article}
\newcommand{\be}{\begin{equation}}
\newcommand{\ee}{\end{equation}}
\newcommand{\bea}{\begin{eqnarray}}
\newcommand{\eea}{\end{eqnarray}}
\newcommand{\ba}{\begin{array}}
\newcommand{\ea}{\end{array}}
\newcommand{\beas}{\begin{eqnarray*}}
\newcommand{\eeas}{\end{eqnarray*}}
\newcommand{\bes}{\begin{equation*}}
\newcommand{\ees}{\end{equation*}}

\begin{document}
\title{\bf Logarithmic curvature correction to Lifshitz and  hyperscaling violation geometries}
\author{J. Sadeghi  $^{a}$\thanks{E-mail: pouriya@ipm.ir}\hspace{1mm}
, M. R. Setare $^{b}$ \thanks{E-mail:
rezakord@ipm.ir}\hspace{1mm},
 S. K. Moayedi $^{c}$\thanks{E-mail: s-moayedi@araku.ac.ir}\hspace{1.5mm}
  F. Pourasadollah$^{a}$\thanks{E-mail: f.pourasadollah@stu.umz.ac.ir}\hspace{1.5mm}\\
$^a$ {\small {\em Sciences Faculty, Department of Physics, Mazandaran University,P.O.Box 47416-95447, Babolsar, Iran }}\\
$^{b}${\small {\em Department of Science, Campus of Bijar,
University of Kurdistan, Bijar, Iran}}\\ $^c$ {\small {\em
Department of Physics, Faculty of Sciences, Arak University, Arak
38156-8-8349, Iran}}\\} \maketitle
\begin{abstract}
 In this note we consider the logarithmic curvature correction to Lifshitz and hyperscaling violation geometries.
 We investigate the effect of this correction to the gauge kinetic function $f(\phi)$ and the effective potential $V(\phi)$.
 For the case of hyperscaling violating we show that the coupling between dilaton and the correction terms exhibit the logarithmic behavior for dilaton
  like for the case of squared curvature correction. We find the unexpected form of gauge kinetic function and potential in the presence of logarithmic terms. We try to resolve the IR singularity of Lifshitz and hyperscaling violation geometries by adding the logarithmic term.  
\end{abstract}
\section{Introduction}
In the recent years, various classes of background in application of condensed matter is
discussed by the holographic method \cite{P1}-\cite{P6}. In that case, the simple generalization is to consider
metric background as dual to scale-invariant field theories and non-conformal invariance.
So, we are going to introduce the metric background as \cite{P7,P8}
\begin{equation}
 ds ^{2} = -\frac{dt^{2}}{r^{2z}}+\frac{(dr^{2}+dx_{i}^{2})}{r^{2}},
\end{equation}
which is invariant under the following scaling,
\begin{equation}
t\rightarrow \lambda^{z}t  \qquad   , \quad x_{i}\rightarrow \lambda
x_{i} \quad ,\qquad r\rightarrow \lambda r.
\end{equation} \\
For the holographic applications to the condensed matter system, it becomes clear that
we need to consider more general classes of metrics than those with the asymptotic AdS
boundaries. Recently, Lifshitz-type theories with hyperscaling violation including
an abelian gauge and scalar dilaton field discussed and the corresponding metric will be as \cite{P9},
\begin{equation}
ds ^{2} = r^{-\frac{2 \theta}{d}}\left[r^{2z}
dt^{2}+\frac{dr^{2}}{r^{2}}+ \sum_{i=1}^{d} r^{2}dx_{i}^{2}\right].
\end{equation}
We note here the metric background includes a dynamical critical
exponent $z$ and a hyperscaling violation exponent $\theta$, also
$d$ is the number of transverse dimensions. The metric  background
(3) is not invariant, but transforms covariantly $ ds\rightarrow
\lambda ^{\theta / d}ds$ which is a defining property of
hyperscaling violation in holographic language. Hyperscaling
violation first mentioned in context of holographic in \cite{P12}. For the
theories with hyperscaling, the entropy behaves as $ S\sim T^{d/z}$.
In this context, the hyperscaling violation exponent $ \theta$ is
related to the transformation of the proper distance, and its
non-invariance implies the violation of hyperscaling of the dual
field theory . Then, the relation between the entropy and
temperature has been modified as $ S\sim T^{(d-\theta)/z}$ . In
general, theory with hyperscaling violation $d-\theta$ plays the
role of an effective space dimensionality for the dual field theory
\cite{P9,P12,P13}. In paper \cite{14} the authors have studied the effect of certain quartic derivative terms on hyperscaling violating solutions, using in particular dimensional reduction techniques which make clear the reinterpretation of $\theta$ as some effective dimensionality. Though they could not achieve any resolution of the curvature singularities using these higher-derivative terms. In \cite{15} have been shown how generalized dimensional reduction techniques can be used to define the hyperscaling violating holographic dictionary, and to what extent things really scale as in $d-\theta$ dimensions.
\\ Also, in order to get a physically sensible dual field
theory, we should construct the null energy condition $(NEC)$ as
$T_{\mu\nu} n^{\mu}n^{\nu}\geq 0$ on the Einstein equation, where
$n^{\mu}n^{\nu}=0$, so the $NEC$ give us,$
(d-\theta)(d(\tau-1)-\theta)\geq0$ and
$(\tau-1)(d+\tau-\theta)\geq0.$
 The Einstein-Maxwell-Dilaton action which is responsible for both Lifshitz and hyperscaling geometries has the following form  \cite{P14,P15,P16},
\begin{equation}
e^{-1}\mathcal{L}= \mathrm{R}-\frac{1}{2}(\partial\phi)^{2}-f(\phi)F_{\mu\nu}F^{\mu\nu}-V(\phi)
\end{equation}
where $f(\phi)=e^{\lambda\phi}$ is defined for both Lifshitz and hyperscaling geometries while the potential follows as,
\begin{eqnarray}
V(\phi) &=& -\Lambda\qquad\qquad\qquad\qquad \mathrm {for\;the\:Lifshitz\:case}, \\
V(\phi) &=& V_{0}e^{\gamma\phi}\qquad\qquad\qquad\, \mathrm {for\:the\:hyperscaling\:case}.
\end{eqnarray}
Both of these spacetime suffer from the null singularity in IR which makes the infrared incomplete \cite{P17}-\cite{P23}. In Refs \cite{P22} and \cite{P23} the authors attempted to resolve the IR singularity by implementing the $R^{2}$ correction to the action (4) for the Lifshitz and hyperscaling violation respectively. \footnote{Recently, in \cite{16}, the authors proposed a model which provides a nonsingular IR completion of hyperscaling violating Lifshitz spacetimes, namely another IR $AdS_4$ rather than $AdS_2\times R^2$. They make use of an effective mass for the vector in the IR to eat up the electric flux and close the $AdS_2\times R^2$ throat. }
 In this paper we are interested  in $f(R)$ theories , in which the Ricci curvature $R$ in the Einstein-Hilbert action of general relativity is replaced by an arbitrary function of $R$ in the Lagrangian \cite{P24}-\cite{P27}. On the more specific case, we would like to investigate the $f(R)$ model with logarithmic term of the form $R+\alpha_{1}R^{2}+\alpha_{2}R^{2}\ln\frac{R}{R_{0}}$.\\
  $f(R)$ theories with logarithmic terms have been discussed in \cite{P25}-\cite{P27}. The removal of singularities in Horava-Lifshitz gravity via higher derivative terms of $R^2$ type have
been discussed in \cite{24}. Recently, inflation in $R^2+R^2 \ln R$ theory was studied in \cite{25}.
  In this work we utilize this correction for the both Lifshitz and hyperscaling violation backgrounds.\\
 This paper is assigned as follows. In section 2, we extend the Einstein-Maxwell-Dilaton system by adding the logarithmic correction and  obtain the resulting
 Lifshitz solution. We repeat the above process for the hyperscaling violation background and try to resolve the IR singularity in section 3 . Finally, in the section 4, we close with the summary.
\section{Lifshitz solutions in the presence of logarithmic correction}
By replacing the $f(R)$ terms, our Einstein-Maxwell-Dilaton action (4) take the following form,
\begin{equation}
S=\int d^{d+2}x \sqrt{-g}\left(R+\Lambda-\frac{1}{2}(\partial\phi)^{2}-f(\phi)F_{\mu\nu}F^{\mu\nu}+\alpha_{1}R^{2}+\alpha_{2}R^{2}\ln(\frac{R}{R_{0}})\right),
\end{equation}
where $R_{0}$ is a positive constant parameter \cite{P26}. Taking the variation of action, we write the Einstein's equations as,
\begin{equation}
E_{\mu\nu}=T_{\mu\nu},
\end{equation}
where,
\begin{eqnarray}
E_{\mu\nu}&=&R_{\mu\nu}-\frac{1}{2}g_{\mu\nu}R-\frac{1}{2}g_{\mu\nu}\left(\alpha_{1}R^{2}+\alpha_{2}R^{2}\ln(\frac{R}{R_{0}}\right)
+2\alpha_{2}\left(g_{\mu\nu}\Box R \ln(\frac{R}{R_{0}})-\nabla_{\mu}\nabla_{\nu}R \ln(\frac{R}{R_{0}})\right)\nonumber\\
&&+\left(2\alpha_{1}+\alpha_{2}\right)\left(g_{\mu\nu}\Box R-\nabla_{\mu}\nabla_{\nu}R\right)+\left(2\alpha_{1}+\alpha_{2}+2\alpha_{2}\ln(\frac{R}{R_{0}})\right)RR_{\mu\nu},
\end{eqnarray}
and
\begin{equation}
T_{\mu\nu}=\frac{1}{2}\partial_{\mu}\phi\partial_{\nu}\phi+2f(\phi)\left(F_{\mu}^{\rho}F_{\nu\rho}-
\frac{1}{4}g_{\mu\nu}F_{\rho\sigma}F^{\rho\sigma}\right)+\frac{1}{2}g_{\mu\nu}\left(\Lambda-\frac{1}{2}\partial^{\rho}\phi\partial_{\rho}\phi\right).
\end{equation}
With these equations, the Maxwell and scalar equations are obtained as,
\begin{equation}
\nabla_{\mu}\left(f(\phi)F^{\mu\nu}\right)=0,
\end{equation}
\begin{equation}
\Box\phi-f'(\phi)F_{\mu\nu}F^{\mu\nu}=0,
\end{equation}
where $f'(\phi)$ is the derivative of $f(\phi)$ with respect to $\phi$. By using the metric ansatz (1), the Maxwell equation (11) leads to
\begin{equation}
F_{rt}=\frac{Q}{f(\phi)}r^{z-d-1},
\end{equation}
where $Q$ is an integration constant. To solve the Einstein's equations, we combine the various component of energy momentum tensor as follows,
\begin{eqnarray}
  T_{t}^{t}-T_{r}^{r} &=& -\frac{1}{2}r^{2}(\partial_{r}\phi)^{2},\nonumber \\
  T_{x}^{x}-T_{t}^{t} &=& 2 Q^{2} f(\phi)^{-1}r^{-2d}, \nonumber\\
  T_{x}^{x}+T_{r}^{r} &=& \Lambda.
\end{eqnarray}
On the other hand one can obtain,
\begin{eqnarray}
  E_{t}^{t}-E_{r}^{r} &=& d(1-z)\left[1-2 K\left(2\alpha_{1}+\alpha_{2}+
  2\alpha_{2}\ln\frac{-2K}{R_{0}}\right)\right],\nonumber \\
  E_{x}^{x}-E_{t}^{t} &=& (z-1)(z+d)\left[1-2K\left(2\alpha_{1}+\alpha_{2}+
  2\alpha_{2}\ln\frac{-2K}{R_{0}}\right)\right],\nonumber\\
  E_{x}^{x}+E_{r}^{r} &=&(z+d)(z+d-1) -2K \alpha_{1}[2z(d-1)+d(d-3)], \nonumber \\
  &&-2K\alpha_{2}\left[-\left(z^{2}+z+2d\right)+[2z(d-1)+d(d-3)]\ln\frac{-2K}{R_{0}}\right],
\end{eqnarray}
where $K=z^{2}+dz+\frac{d(d+1)}{2}$. By combining the above equations, the final solution in the presence of logarithmic correction is given by
the Lifshitz metric (1), the Maxwell field (13), and the dilaton
\begin{equation}
\phi=\beta\ln r+\phi_{0},
\end{equation}
where the parameter of the solution, $\beta$, charge and the cosmological constant $\Lambda$ are given in the terms of $z$ and $\alpha_{1}$ and $\alpha_{2}$ according to,
\begin{equation}
\beta^{2}=2d(z-1)\left[1-2 K\left(2\alpha_{1}+\alpha_{2}+
  2\alpha_{2}\ln\frac{-2K}{R_{0}}\right)\right],
\end{equation}
\begin{equation}
Q^{2}e^{-\lambda\phi_{0}}=\frac{(z-1)(z+d)}{2}\left[1-2K\left(2\alpha_{1}+\alpha_{2}+
  2\alpha_{2}\ln\frac{-2K}{R_{0}}\right)\right],
\end{equation}
and
\begin{eqnarray}
\Lambda&=&(z+d)(z+d-1) -2K \alpha_{1}[2z(d-1)+d(d-3)]\nonumber\\
  &&-2K\alpha_{2}\left[-\left(z^{2}+z+2d\right)+[2z(d-1)+d(d-3)]\ln\frac{-2K}{R_{0}}\right].
\end{eqnarray}
By replacing the $f(R)$ terms, we see that the  action (7) admits Lifshitz solutions with an electric background gauge potential and $\phi\propto \ln r$. The effect of extra terms in the action is to renormalize the cosmological constant and electric charge by inducing correction in the terms of  $\alpha_{1}$, $\alpha_{2}$ and $z$.
\section{Logarithmic correction to the hyperscaling violating background}
In this section we generalize the solution of last section to non-zero $\theta$ and investigate the effects of logarithmic correction on the solutions of
hyperscaling violating backgrounds. For this purpose we introduce the hyperscaling violating action in the presence of $f(R)$ terms as,
\begin{equation}
S=\int d^{d+2}x \sqrt{-g}\left(R+V(\phi)-\frac{1}{2}(\partial\phi)^{2}-f(\phi)F_{\mu\nu}F^{\mu\nu}+g(\phi)(\alpha_{1}R^{2}+\alpha_{2}R^{2}\ln(\frac{R}{R_{0}}))\right).
\end{equation}
where the $g(\phi)$ function added to derive the hyperscaling violating solution [21],\cite{P28}. The equations of motion of the above action are obtained as follows,
\begin{eqnarray}
E_{\mu\nu}&=&R_{\mu\nu}-\frac{1}{2}g_{\mu\nu}R +g(\phi)\left[-\frac{1}{2}g_{\mu\nu}\left(\alpha_{1}R^{2}+\alpha_{2}R^{2}\ln(\frac{R}{R_{0}})\right)
+\left(2\alpha_{1}+\alpha_{2}+2\alpha_{2}\ln(\frac{R}{R_{0}})\right)RR_{\mu\nu}\right]\nonumber\\
&&+2\alpha_{2}\left(g_{\mu\nu}\Box g(\phi) R \ln(\frac{R}{R_{0}})-\nabla_{\mu}\nabla_{\nu}g(\phi)R \ln(\frac{R}{R_{0}})\right)+\left(2\alpha_{1}+\alpha_{2}\right)g_{\mu\nu}\Box g(\phi)R\nonumber\\
&&-\left(2\alpha_{1}+\alpha_{2}\right)\nabla_{\mu}\nabla_{\nu}g(\phi)R,
\end{eqnarray}
and
\begin{equation}
\Box\phi-f'(\phi)F_{\mu\nu}F^{\mu\nu}+g'(\phi)\left(\alpha_{1}R^{2}+\alpha_{2}R^{2}\ln(\frac{R}{R_{0}})\right)=-\frac{dV(\phi)}{d\phi}.
\end{equation}
Lets us take the general form of $V(\phi)$ and $f(\phi)$ in the above equations of motion and $g(\phi)=e^{\eta\Phi}$ with $\Phi=\Phi_{0}+\omega\ln r$ . By implementing the metric anstaz (3) and repeating the progress of last section we have,
\begin{eqnarray}
  -\frac{1}{2}r^{2+2\frac{\theta}{d}}(\partial\phi)^{2} &=& r^{\frac{2\theta}{d}}\left[(d-\theta)(1-z+\frac{\theta}{d})\Big[1-e^{\eta\Phi_{0}} P\Big(2\alpha_{1}+\alpha_{2}+
  2\alpha_{2}\ln\frac{-Pr^{\frac{2\theta}{d}}}{R_{0}}\Big)\Big]+\frac{4\alpha_{2}\theta}{d}Pe^{\eta\Phi_{0}}\right],\nonumber \\
 2Q^{2}r^{2(\theta-d)}f(\phi)^{-1} &=& (z-1)(z+d-\theta)r^{\frac{2\theta}{d}}\left[1-e^{\eta\Phi_{0}}P\Big(2\alpha_{1}+\alpha_{2}+
  2\alpha_{2}\ln\frac{-r^{\frac{2\theta}{d}}P}{R_{0}}\Big)\right],\nonumber\\
  V(\phi)&=&(z+d-\theta)(z+d-\theta-1)r^{\frac{2\theta}{d}} -r^{\frac{2\theta}{d}}e^{\eta\Phi_{0}}P^{2}\left[\alpha_{1} +\alpha_{2}\ln\frac{-r^{\frac{2\theta}{d}}P}{R_{0}}\right]\nonumber \\
  &&Pr^{\frac{2\theta}{d}}e^{\eta\Phi_{0}}\left[z\Big(z+1-2\frac{\theta}{d}\Big)+(d-\theta)(2-\frac{\theta}{d})\right]\left[2\alpha_{1}+\alpha_{2}+
  2\alpha_{2}\ln\frac{-Pr^{\frac{2\theta}{d}}}{R_{0}}\right]\nonumber \\
  &&+4\alpha_{2}Pr^{\frac{2\theta}{d}}e^{\eta\Phi_{0}}\Big(2d-2\theta+2z-\frac{\theta}{d}\Big),
\end{eqnarray}
where $P=(d-\theta)\left((d+1)(1-\frac{\theta}{d})+2z\right)+2z(z-\frac{\theta}{d})$. The first of the above equations shows that,
\begin{equation}
\phi=\frac{2}{3\omega}\left[\Phi_{0}+\omega\ln r\right]^{\frac{3}{2}},
\end{equation}
with $\Phi_{0}=2(\theta -d)(1-z+\frac{\theta}{d})\left[1-e^{\eta\Phi_{0}} P\left(2\alpha_{1}+\alpha_{2}+
  2\alpha_{2}\ln\frac{-P}{R_{0}}\right)\right]-8\alpha_{2}\frac{\theta}{d}Pe^{\eta\Phi_{0}}$ and $\omega=\frac{-2\theta}{\eta d}=8e^{\eta\Phi_{0}} P(d-\theta)(1-z+\frac{\theta}{d})\frac{\theta}{d}\alpha_{2}$ and we scale $\Phi=(\frac{3\omega\phi}{2})^{\frac{2}{3}}$. By investigating the relevant relations to the potential and the charge, one can find that the exponential form for $V(\phi)$ and $f(\phi)$ is not enough. For solving this problem we define $f(\phi)^{-1}=e^{-\lambda \Phi}(\Phi-\lambda_{0})$ and $V(\phi)=e^{\gamma \Phi}(\Phi-\gamma_{0})$. With these constructions one can easily check the following values for $\lambda$ and $\gamma$,
  \begin{equation}
  \lambda=\frac{2}{\omega}(\theta-d-\frac{\theta}{d}),\qquad\qquad \gamma=-\eta=\frac{-2\theta}{d\omega}\:.
  \end{equation}
  By substituting the above relations in the equation (23) we obtain,
  \begin{eqnarray}
  2Q^{2}&=& \frac{e^{(\eta+\lambda)\Phi_{0}}(1-z)(d-\theta+z)}{4(d-\theta)(1+\frac{\theta}{d}-z)},\nonumber \\
   \lambda_{0} &=& \Phi_{0}+2(d-\theta)(1+\frac{\theta}{d}-z)e^{-\eta\Phi_{0}}\left[1-e^{\eta\Phi_{0}}P\left(2\alpha_{1}+\alpha_{2}+
  2\alpha_{2}\ln\frac{-P}{R_{0}}\right)\right],
      \end{eqnarray}
and
\begin{eqnarray}
V_{0}&=&-\frac{(d-\theta+z)(d-\theta+z-\frac{\theta}{d})+d-\theta+z(z-\frac{\theta}{d})}{4(d-\theta)
  (1+\frac{\theta}{d}-z)}e^{-2\gamma\Phi_{0}},\nonumber\\
    \gamma_{0} &=& \Phi_{0}-\frac{4e^{\gamma\Phi_{0}}(d-\theta)
  (1+\frac{\theta}{d}-z)}{(d-\theta+z)(d-\theta+z-\frac{\theta}{d})+d-\theta+z(z-\frac{\theta}{d})}\Big[(d-\theta+z)(d-\theta+z-1)\nonumber\\
&&+ e^{\eta\Phi_{0}}\Big\{
  \Big((d-\theta)(2-\frac{\theta}{d})+z(1+z-2\frac{\theta}{d})\Big)P(2\alpha_{1}+\alpha_{2}+
  2\alpha_{2}\ln\frac{-P}{R_{0}})\nonumber\\
 &&P^{2}(\alpha_{1}+\alpha_{2}\ln\frac{-P}{R_{0}})-\frac{4\alpha_{2}}{d}P(1+2d-2\theta+2z)\Big\}\Big].
\end{eqnarray}
We see that the logarithmic correction to the hyperscaling action, don't lead to a solution with the pervious expected form of $V(\phi)$ and $f(\phi)$.
As we showed in this section, to have a correct solution we have to change the form of potential and gauge kinetic function. We expect that our results to be in agreement with the results of previous section in the limit of $\theta=0$ for the Lifshitz geometry. This statement can be obtained when we use the following equalities,
\begin{eqnarray}
 2Q^{2}e^{-\lambda \Phi_{0}}(\Phi_{0}-\lambda_{0})\Big|_{\theta\rightarrow0\;(hyperscaling)}&=&2Q^{2}e^{-\lambda \Phi_{0}}\Big|_{Lifshitz}\nonumber\\
 V_{0}e^{\gamma \Phi_{0}}(\Phi_{0}-\gamma_{0})\Big|_{\theta\rightarrow0\:(hyperscaling)}&=&V_{0}e^{\gamma \Phi_{0}}\Big|_{Lifshitz},
\end{eqnarray}
so, the gauge kinetic function and the potential receive to the expected exponential form. One can also check the $\alpha_{2}\ll\alpha_{1}$ condition which gives the $R^{2}$ gravity. In this statement $f(\phi)$ and $V(\phi)$ functions receive to the results of \cite{P22} and \cite{P23} for Lifshitz and hyperscaling violating geometries by using the relation (28).
\paragraph{ Resolving the singularity}
 We follow the Ref \cite{P23} to investigate the behavior of hyperscaling violating solutions in the IR in the presence of logarithmic correction. For this purpose we work in the (a-b) gauge and so take the following metric ansatz,
\begin{equation}
ds^{2}=a^{2}(r)\Big(-dt^{2}+dr^{2}+b^{2}(r)(dx^{2}+dy^{2})\Big).
\end{equation}
By changing the variable in the above relation in a way that $z\tilde{r}=r^{-z}$, $\tilde{x}=z^{1-\frac{1}{z}}x$ we receive,
\begin{equation}
ds^{2}=\tilde{L}^{2}\tilde{r}^{\frac{\theta}{z}-2}\Big(-dt^{2}+d\tilde{r}^{2}+\tilde{r}^{2-\frac{2}{z}}d\tilde{x}^{2}\Big),
\end{equation}
where $\tilde{L}^{2}=L^{2}z^{\frac{\theta}{z}-2}$ and the hyperscaling violating solution in this gauge is,
\begin{equation}
ds^{2}=\tilde{L}^{2}r^{\theta-\theta\tilde{z}-2}\Big(-dt^{2}+dr^{2}+r^{2\tilde{z}}dx^{2}\Big)\qquad \mathrm{with} \qquad z=\frac{1}{1-\tilde{z}}.
\end{equation}
The field parametrization in the (a-b) gauge becomes,
\begin{equation}
F_{rt}=\frac{Q}{f(\phi)b^{2}}.
\end{equation}
To repeat the process of last section in this gauge we need the following relations,
\begin{eqnarray}
  R_{t}^{t}-R_{r}^{r}&=&\frac{1}{a^{2}} \Big(\frac{2b''}{b}+\frac{a''}{a}-\frac{4a'b'}{a b}\Big),\nonumber\\
  R_{x}^{x}-R_{t}^{t}&=&-\frac{1}{a^{2}} \Big(\frac{b''}{b}+\frac{2a'b'}{a b}+\frac{b'^{2}}{b^{2}}\Big),\\
  R_{x}^{x}+R_{r}^{r}&=&-\frac{1}{a^{2}} \Big(3\frac{b''}{b}+4\frac{a''}{a}+6\frac{a'b'}{a b}-2\frac{a'^{2}}{a^{2}}+\frac{b'^{2}}{b^{2}}\Big),\nonumber\\
  R&=&-\frac{1}{a^{2}}\Big(\frac{2b''}{b}+\frac{3a''}{a}+\frac{6a'b'}{a b}+\frac{b'^{2}}{b^{2}}\Big).\nonumber
  \end{eqnarray}
  If we consider an initial solution of $AdS_{2}\times R^{2}$ in the IR as,
  \begin{equation}
  a(r)=\frac{1}{r},\qquad b(r)=b_{I}r,\qquad\phi(r)=\phi_{I},
  \end{equation}
  and define $g(\phi)=\frac{1}{2}(c_{1}e^{\eta\phi}+c_{2})$, then by using the relations (33) we obtain,
  \begin{equation}
  V(\phi_{I})=1+2\alpha_{2}g(\phi_{I}),\qquad\qquad \frac{Q^{2}}{b_{I}^{4}}=\frac{f(\phi_{I})}{2}\Big(1-2g(\phi_{I})(2\alpha_{1}+\alpha_{2}+2\alpha_{2}\ln\frac{-2}{R_{0}})\Big),
  \end{equation}
  and the scalar equation becomes,
  \begin{equation}
  \frac{f'(\phi_{I})}{f(\phi_{I})}\Big(1-2g(\phi_{I})(2\alpha_{1}+\alpha_{2}+2\alpha_{2}\ln\frac{-2}{R_{0}})\Big)+4g'(\phi_{I})( \alpha_{1}+\alpha_{2}\ln\frac{-2}{R_{0}})+V'(\phi_{I})=0.
  \end{equation}
  Now by considering $f(\phi)=e^{\lambda\phi}$, $V(\phi)=V_{0}e^{\gamma\phi}$, we have the following solution in the IR
  \begin{eqnarray}
     V_{0}e^{\gamma\phi_{I}} &=& 1+\alpha_{2}(c_{1}e^{\eta\phi_{I}}+c_{2}),\nonumber \\
    \phi_{I} &=& \frac{1}{\eta}\ln \frac{c_{2}\Big(2\lambda(\alpha_{1}+\alpha_{2}\ln\frac{-2}{R_{0}})+\alpha_{2}(\lambda-\gamma)\Big)-(\lambda+\gamma)}
    {c_{1}\Big(2(\eta-\lambda)(\alpha_{1}+\alpha_{2}\ln\frac{-2}{R_{0}})+\alpha_{2}(\gamma-\lambda)\Big)},\\
    \frac{Q^{2}}{b_{I}^{4}}&=&\frac{e^{\lambda\phi_{I}}}{2}\Big(1-(2\alpha_{1}+\alpha_{2}+2\alpha_{2}\ln\frac{-2}{R_{0}})
    \frac{c_{2}\Big(2\eta(\alpha_{1}+\alpha_{2}\ln\frac{-2}{R_{0}})\Big)-(\lambda+\gamma)}
    {2(\eta-\lambda)(\alpha_{1}+\alpha_{2}\ln\frac{-2}{R_{0}})+(\gamma-\lambda)}\Big).\nonumber
  \end{eqnarray}
  As we see, the behavior of potential and kinetic gauge potential is exponential in the IR. The above solution only makes sense for a certain value of $\lambda$, $\gamma$, $\alpha_{1}$, $\alpha_{2}$, $c_{1}$ and $c_{2}$. We don't discuss about the numerical space of parameters in this paper as \cite{P24}and \cite{P25} and leave this work for future.\\
  Resolving the IR singularity for the Lifshitz geometry with the logarithmic correction can be done by repeating the above process as in \cite{P22}.
  In this case, we should replace the potential with the cosmological constant. This statement leads to introducing $g(\phi)=C=\frac{1}{2(2\alpha_{1}+\alpha_{2}+2\alpha_{2}\ln\frac{-2}{R_{0}})}$. Now by considering $f(\phi)=e^{\lambda\phi}$, we receive to the pure gravitational solution in which $Q^{2}=0$ and the cosmological constant is $\Lambda=1+\frac{\alpha_{2}}{2\alpha_{1}+\alpha_{2}+\alpha_{2}\ln\frac{-2}{R_{0}}}$.
  If we want to achieve a solution with non-zero charge we have to change the definition of $f(\phi)$ and equal it to a constant value.
 \section{Summary}
In this paper, we studied the logarithmic correction to the Lifshitz and hyperscaling geometries. For the Lifshitz background we found that this correction effects on the charge and the cosmological constant and so these parameters have a new definition in the terms of dynamical exponent and the constants of correction. We obtained the dilaton through the logarithmic dependance on the radial coordinate and found that the form of gauge kinetic function exponentially remains unchanged. In the hyperscaling violating background the consequences of the logarithmic correction are more. We showed that to have a correct solution in the presence of this correction we have to change the form of potential, gauge kinetic function and the dilaton function. These functions have a new definition in terms of $\theta$ , $z$, $\alpha_{1}$ and $\alpha_{2}$. We checked that our results for the potential and the gauge kinetic function in the hyperscaling violating case in the limit of $\theta=0$ are in agreement with what we obtained in the Lifshitz case. At the end of our work in this paper we studied the IR behavior of hyperscaling violation and Lifshitz solutions in the presence of logarithmic term. In this limit for the hyperscaling violating geometry, we found that  $V(\phi)$ and $f(\phi)$ get back to their exponential forms and so we couldn't find a solution with the new form of these functions. For the Lifshitz case, we received to the pure gravitational result where the matter field has been decoupled and found that to have a solution with non-zero charge we should change the form of kinetic function as the hyperscaling violating case.\\
Recently the authors of \cite{26} have provided a further motivation for embedding Lifshitz
geometries with hyperscaling violation in higher-order theories like
our, namely, the fact that including such terms modifies the
structure of divergences of the holographic entanglement entropy
formula, in some cases producing new universal terms. Actually, one
might wonder what kind of terms would be produced by the
$R^2+\ln f(R)$ theory we considered in this paper.
\section{Acknowledgments}
We thank S. Odintsov and  P. Bueno for bringing our attention to interesting papers Refs.[27]-[29] and Ref.[33] respectively. We also thank B. Gouteraux, for his comment and bringing our attention to the papers [12,13, 24].


\begin{thebibliography}{11}
\bibitem{P1}
L. Huijse, S. Sachdev and B. Swingle, "Hidden Fermi surfaces in compressible states of
gauge-gravity duality," Physical Review B 85, 035121 (2012)[arXiv:1112.0573 [cond-mat.str-el]].
\bibitem{P2}
C. Charmousis, B. Gouteraux, B. S. Kim, E. Kiritsis and R. Meyer, "Effective Holographic Theories for low-temperature condensed matter systems," JHEP 1011, 151 (2010) [arXiv:1005.4690
[hep-th]].
\bibitem{P3}
N. Ogawa, T. Takayanagi and T. Ugajin, "Holographic Fermi Surfaces and
Entanglement Entropy," JHEP01, 125 (2012) [arXiv:1111.1023 [hep-th]].
\bibitem{P4}
K. Goldstein, N. Iizuka, S. Kachru, S. Prakash, S. P. Trivedi and A. Westphal,
"Holography of Dyonic Dilaton Black Branes," JHEP 1010, 027 (2010)
[arXiv:1007.2490 [hep-th]].
\bibitem{P5}
S. Sachdev, "A model of a Fermi liquid using gauge-gravity duality", Physical Review D 84, 066009 (2011).
\bibitem{P6}
S. A. Hartnoll, J. Polchinski, E. Silverstein, D. Tong, "Towards strange metallic holography," JHEP
1004:120, (2010), [arXiv:0912.1061 [hep-th]].
\bibitem{P7}
S. Kachru, X. Liu and M. Mulligan, "Gravity Duals of Lifshitz-like Fixed Points", Phys.
Rev. D 78, 106005 (2008).
\bibitem{P8}
M. Taylor, Non-relativistic holography, arXiv:0812.0530 [hep-th].
\bibitem{P9}
X. Dong, S. Harrison, S. Kachru, G. Torroba and H.Wang, "Aspects of holography for theories
with hyperscaling violation," JHEP 1206, 041 (2012) [arXiv:1201.1905 [hep-th]].
\bibitem{P12}
B. Gouteraux and E. Kiritsis, "Generalized Holograghic Quantum Critiality at Finite Density", JHEP 1112, 036 (2011).
\bibitem{P13}
B. S. Kim, "Schr\"{o}dinger Holography with and without Hyperscaling Violation", JHEP 06, 116 (2012).
\bibitem{14}C. Charmousis, B. Goutéraux, E. Kiritsis, "Higher-derivative scalar-vector-tensor theories: black holes, Galileons, singularity cloaking and holography", JHEP 1209, 011, (2012)  [arXiv:1206.1499 [hep-th]].
\bibitem{15}B. Goutéraux, J. Smolic, M. Smolic, K. Skenderis, M. Taylor, "Holography for Einstein-Maxwell-dilaton theories from generalized dimensional reduction", JHEP 1201, 089, (2012) 	[arXiv:1110.2320 [hep-th]].
\bibitem{P14}
K. Goldstein, S. Kachru, S. Prakash and S. P. Trivedi, "Holography of Charged Dilaton Black
Holes", JHEP 1008, 078 (2010) [arXiv:0911.3586 [hep-th]].
\bibitem{P15}
M. Cadoni, G. D'Appollonio and P. Pani, "Phase Transitions Between Reissner-Nordstrom and
Dilatonic Black Holes in 4D AdS Spacetime", JHEP 1003, 100, (2010) [arXiv:0912.3520 [hep-th]].
\bibitem{P16}
 C.-M. Chen and D.-W. Pang, "Holography of Charged Dilaton Black Holes in General
Dimensions", JHEP 1006 (2010) 093 [arXiv:1003.5064 [hep-th]].
\bibitem{P17}
G. T. Horowitz, B. Way, "Lifshitz Singularities" [arXiv:1111.243[hep-th]].
\bibitem{P18}
J. Bhattacharya, S. Cremonini, A. Sinkovics, "On the IR completion of geometries with hyperscaling violation"
[arXiv:1208.1752 [hep-th]].
\bibitem{P19}
Y. Lei, S. F. Ross, "Extending the nonsingular hyperscaling violeting space times" [arXiv:1310.5878 [hep-th]].
\bibitem{P20}
K.Copsey, R.Mann "Singularities in Hyperscaling Violation" [arXiv:1210.1231v2 [hep-th]].
\bibitem{P21}
S. Harrison, S. Kachru and H. Wang, "Resolving Lifshitz Horizons" [arXiv:1202.6635[hep-th]].
\bibitem{P22}
G. Knodel, J. T. Liu, "Higher derivative corrections to Lifshitz backgrounds", [arXiv:1305.3279 [hep-th]].
\bibitem{P23}
M. Ghodrati, "Hyperscaling Violating Solution in Coupled Dilaton-Squared Curvature Gravity", [arXiv:1404.5399 [hep-th]].
\bibitem{16}J. Bhattacharya, S. Cremonini, B. Goutéraux, "Intermediate scalings in holographic RG flows and conductivities", [arXiv:1409.4797 [hep-th]].
\bibitem{P24}
J. Q. Guo and A. V. Frolov, "Cosmological dynamics in f(R) gravity", Phys. Rev. D 88, 124036 (2013) [arXiv:1305.7290[hep-th]].
\bibitem{P25}
J. He, B. Li and Y. Jing, "Revisiting the matter power spectra in f(R) gravity", Phys. Rev. D 88, 103507 (2013) [arXiv:1305.7333[hep-th]].
\bibitem{23}S. Nojiri, S. D. Odintsov, "Modified gravity with $\ln R$ terms and cosmic acceleration",  Gen. Rel. Grav. 36, 1765, (2004),  	 [arXiv:hep-th/0308176]; S. Nojiri, S. D. Odintsov, "Introduction to Modified Gravity and Gravitational Alternative for Dark Energy", Int. J. Geom. Meth. Mod. Phys. 4, 115, (2007),  [arXiv:hep-th/0601213].
 \bibitem{24}S. Carloni, M. Chaichian, S. Nojiri, S. D. Odintsov, M. Oksanen, A. Tureanu, "Modified first-order Horava-Lifshitz gravity: Hamiltonian analysis of the general theory and accelerating FRW cosmology in power-law $F(R)$ model" Phys. Rev. D82, 065020, (2010), [arXiv:1003.3925 [hep-th]].
 \bibitem{25}K. Bamba, G. Cognola, S. D. Odintsov, S. Zerbini, "One-loop Modified Gravity in de Sitter Universe, Quantum Corrected Inflation, and its Confrontation with the Planck Result" Phys. Rev. D 90, 023525 (2014), [arXiv:1404.4311 [gr-qc]].

\bibitem{P26}
H. Alavirad and J. M. Well, "Modified gravity with logarithmic curvature corrections and the structure of
relativistic stars", Phys. Rev. D 88, 124034 (2013) [arXiv:1307.7977[hep-th]].
\bibitem{P27}
M. B. Baibosunov, V. T. Gurovich, and U. M. Imanaliev, ""Model of the early universe in f(R) theory", Soviet Journal of Experimental and
Theoretical Physics 71, 636 (1990); A. Vilenkin, Phys. Rev. D 32, 2511 (1985); G. M. Shore,
Annals of Physics 128, 376 (1980).
\bibitem{P28}
E. Shaghoulian, "FRW cosmologies and hyperscaling-violating geometries: higher curvature corrections, ultrametricity,
Q-space/QFT duality, and a little string theory," [arXiv:1308.1095 [hep-th]].
\bibitem{26}P. Bueno, P. F. Ramirez, "Higher-curvature corrections to holographic entanglement entropy in geometries with hyperscaling violation", 	 [arXiv:1408.6380 [hep-th]]
\end{thebibliography}
\end{document}